\documentclass[prb,twocolumn,showpacs,preprintnumbers,amsmath,amssymb]{revtex4}
\usepackage{graphicx}
\usepackage{dcolumn}
\usepackage{bm}
\usepackage{trfsigns}
\newcommand \be{\begin{eqnarray}}
\newcommand \ee{\end{eqnarray}}
\newcommand \ba{\begin{align}}

\newcommand \V{\vec}

\begin{document}
\title{Correlational latent heat by nonlocal quantum kinetic theory}
\author{K. Morawetz$^{1,2,3}$
}
\affiliation{$^1$M\"unster University of Applied Sciences,
Stegerwaldstrasse 39, 48565 Steinfurt, Germany}
\affiliation{$^2$International Institute of Physics- UFRN,
Campus Universit\'ario Lagoa nova,
59078-970 Natal, Brazil}
\affiliation{$^{3}$ Max-Planck-Institute for the Physics of Complex Systems, 01187 Dresden, Germany
}
\begin{abstract}
 The kinetic equation of nonlocal and non-instantaneous 
character unifies the achievements of the 
transport in dense quantum gases with the Landau theory of quasiclassical transport in Fermi systems. Large cancellations 
in the off-shell motion appear which are hidden usually 
in non-Markovian behaviors. The remaining corrections
are expressed in terms of shifts in space and time that characterize
the non-locality of the scattering process. In this way quantum
transport is possible to recast into a quasi-classical picture. The balance equations for the density, momentum,
energy and entropy include besides quasiparticle also the correlated two-particle contributions beyond the Landau theory.  The medium effects on binary collisions are shown to
mediate the latent heat, i.e., an energy conversion between correlation
and thermal energy. For Maxwellian particles with time-dependent s-wave scattering, the correlated parts of the observables are calculated and a sign change of the latent heat is reported at a universal ratio of scattering length to the thermal De Broglie wavelength. This is interpreted as a change from correlational heating to cooling.
\end{abstract}

\maketitle

\section{Overview  about nonlocal kinetic theory}

\subsection{Introduction}

To
extend the validity of the
Boltzmann equation to moderately dense gases, Clausius and Boltzmann included
the space non-locality of binary collisions \cite{CC90}. After
one century, virial corrections won new interest as they can
be incorporated into Monte Carlo simulation methods \cite{AGA95}.
The microscopic theory of nonlocal corrections to the collision integral 
has been pioneered within the theory of gases by many authors
\cite{W60,S64,B69,TS70,SS71,RS76,B75,M89,La89,NTL89,L90a,H91,LM90,NTL91,NTL91a,S95,SML95}.

In this paper a correlational latent heat in strongly coupled nonequilibrium systems is reported based on the development of a nonlocal extension of the Boltzmann equation. The latter one is reviewed in the next three subchapters of the introduction in order to make the reader acquainted with the consistent extensions of quantum Boltzmann equation and nonequilibrium thermodynamics beyond the Landau theory. The new result is then reported in chapter \ref{LowenergyT} as an actual analytical calculation of all corrections to the quantum Boltzmann equation for a model of time-dependent s-wave scattering. This highlights the nonlocal corrections derived earlier and might be of importance to nonequilibrium quantum systems in cold atoms, plasma physics, and nuclear physics.

\subsection{Extended quasiparticle picture}
In the limit of small scattering
rates, the transport equation for the Green's function is converted
into the kinetic equation of Boltzmann type by the extended quasiparticle
approximation derived for small
scattering rates \cite{SLM96,MLS00,LSM97}. One introduces an effective quasiparticle
distribution $f$ from which the Wigner distribution $\rho$ can be
constructed
\begin{equation}
\rho[f]=f+\int{d\omega\over 2\pi} {\wp\over\omega-\varepsilon}
{\partial\over\partial\omega}
\left((1-f)\sigma^<_\omega-f\sigma^>_\omega\right).
\label{tr2a}
\end{equation}
Here $\sigma^{>}$ and $\sigma^{<}$ denote the selfenergies describing all correlations and $\varepsilon$ is the quasiparticle energy.  
The limit of small scattering rates has been first introduced by Craig
\cite{C66a} and an inverse relation $f[\rho]$ had been constructed
\cite{BD68}. For equilibrium non-ideal plasmas this approximation has been 
employed by \cite{SZ79,KKL84} and has been used under the name of the generalized
Beth-Uhlenbeck approach by \cite{SR87} in nuclear matter 
for studies of the correlated density.  The name "extended quasiparticle 
approximation" finally has been used for the study of the mean removal energy and high-momenta 
tails of Wigner's distribution \cite{KM93}. 
The non-equilibrium form has been derived as the modified Kadanoff and Baym ansatz \cite{SL95}.

\subsection{Nonlocal kinetic equation}
The resulting quantum kinetic theory unifies the achievements of transport in
dense gases with the
quantum transport of dense Fermi systems \cite{SLMa96,SLM96,LSM97,MLS00}. The
quasiparticle drift of Landau's
equation, 
 \begin{equation}
{\partial f_1\over\partial t}+{\partial\varepsilon_1\over\partial \V{k}}
{\partial f_1\over\partial \V{r}}-{\partial\varepsilon_1\over\partial \V{r}}
{\partial f_1\over\partial \V{k}}
=I_1^{\mathrm{in}}-I_1^{\mathrm{out}}
\label{vc17}
\end{equation}
is connected with a dissipation governed by a nonlocal and non-instant
scattering integral in the spirit of Enskog corrections. For the scattering-out (lost term) it reads
\ba
&I_1^{\mathrm{out}}=
\sum_b\int {d^3 \! p\over(2\pi)^3}{d^3 \! q\over(2\pi)^3}
2\pi\delta\left(\varepsilon_1+\varepsilon_2-\varepsilon_3
-\varepsilon_4+2\Delta_E\right)
\nonumber\\
&\times
\Biggl(1+{1\over 2}{\partial\V \Delta_2\over\partial \V r}
+{\partial\varepsilon_2\over\partial r}
{\partial\V \Delta_2\over\partial\omega}\Biggr)_{\omega=\varepsilon_1+\varepsilon_2}
f_1f_2(1-f_3-f_4)
\nonumber\\
&\times
\left|{\cal T}\!\Bigl(\!\varepsilon_1\!+\!\varepsilon_2\!+\!\Delta_E,
\V k\!+\!{\V \Delta_K\over 2},\!\V p+\!{\V \Delta_K\over 2},\V q,\V r\!+\!{\V \Delta_r},
t\!+\!{\Delta_t\over 2}\!\Bigr)\!\right|^2.
\label{Iout}
\end{align}
and the scattering-in (gain term) by replacing $f\leftrightarrow 1-f$ and changing the signs of the shifts.
The distribution functions and observables have the arguments
\begin{equation}
\begin{array}{rcl}
\varepsilon_1&\equiv&\varepsilon_a(\V k,\V r,t),\\
\varepsilon_2&\equiv&\varepsilon_b(\V p,\V r+\V \Delta_2,t),\\
\varepsilon_3&\equiv&\varepsilon_a(\V k-\V q+\V \Delta_K,\V r+\V \Delta_3,t+\Delta_t),\\
\varepsilon_4&\equiv&\varepsilon_b(\V p+\V q+\V \Delta_K,\V r+\V \Delta_4,t+\Delta_t),
\end{array}
\label{vc9dinv}
\end{equation}
with the transferred momentum $\V q$. 

In the scattering-out (scattering-in is analogous) one can see the
distributions of quasiparticles $f_1f_2$ describing the probability of a given
initial state characterized by momenta $k$ and $p$ for the binary collision partners. The hole distributions $(1-f)$ is the
probability that the requested final states are empty. Both combine together in the final-state occupation factors $(1-f_3)(1-f_4)+f_3f_4=1-f_3-f_4$.
The scattering rate covers the energy-conserving
$\delta$-function, and the differential cross section is given by the
modulus of the T-matrix, $|{\cal T}|$, reduced by the wave-function renormalizations 
\cite{K95} which are higher order in small scattering rates.

The corrections to the quantum Boltzmann equation
are expressed in terms of shifts in space and time that characterize the
non-locality of the scattering process \cite{MLSK99}. 
These $\Delta$'s are derivatives of the scattering phase shift $\phi$ of the T-matrix
${\cal T}_R=\left|{\cal T}\right|{\rm e}^{i\phi}$,
according to the following list
\ba
&{\V \Delta_K}={1\over 2}{\partial\phi\over\partial \V r},
&\Delta_E&=-{1\over 2}{\partial\phi\over\partial t},
\qquad \Delta_t={\partial\phi\over\partial\omega},
\nonumber\\
&\V \Delta_2=
{\partial\phi\over\partial \V p}
-{\partial\phi\over\partial \V q}
-{\partial\phi\over\partial \V k},
&\V \Delta_3&=-{\partial\phi\over\partial \V k},
\nonumber\\
&{\V \Delta_4}=-{\partial\phi\over\partial \V q}-{\partial\phi\over\partial \V k},
&{\V \Delta_r}&={1\over 4}\left(\V \Delta_2\!+\!\V \Delta_3\!+\!{\V \Delta_4}\right)
.
\label{vc4g}
\end{align}

As special limits, this kinetic theory includes the Landau theory as well as the Beth-Uhlenbeck equation of state \cite{SRS90,MR94} which means correlated pairs.
The medium effects on binary collisions are shown to
mediate the latent heat which is the energy conversion between correlation
and thermal energy \cite{LSM99,LSM97}. In this respect the seemingly contradiction between particle-hole symmetry and time reversal symmetry in the collision integral was solved \cite{SLM98}.
Compared to the Boltzmann-equation, the presented form of virial
corrections only slightly increases the numerical demands in
implementations \cite{MSLKKN99,MT00,MLNCCT01} since large cancellations
in the off-shell motion appear which are hidden usually
in non-Markovian behaviors. Details how to implement the nonlocal kinetic equation into existing Boltzmann codes can be found in \cite{MLNCCT01}.
In this way quantum
transport is possible to recast into a quasiclassical picture suited for simulations.

\subsection{Balance equations}
The balance equations for the density, momentum,
energy and entropy include quasiparticle contributions and the correlated two-particle contributions beyond the Landau theory. A number of attempts have been made to modify the 
Boltzmann equation so that its equilibrium limit would cover at 
least the second 
virial coefficient \cite{B69,Ba84,S91}. The corrections to 
the Boltzmann equation have the form of gradients or nonlocal contributions 
to the scattering integral. The nature of two-particle correlations induces gradients and therefore nonlocal kinetic and exchange energies \cite{MS91,Mar97}.

We multiply
the kinetic equation \eqref{vc17} with a variable $\xi_1=1,\V {k},\varepsilon_1,-k_{\mathrm B}\ln[f_1/(1-f_1)]$
and integrate over momentum. It results in the equation of continuity, the
Navier-Stokes equation, the energy balance and the evolution of the
entropy, respectively. 
All these conservation laws or balance equations for the mean thermodynamic observables have the form \cite{M17,M17b} 
\be
{\partial \langle \xi^{\rm qp}+\xi^{\rm mol} \rangle \over \partial t}+{\partial (\V j^{\rm qp}_\xi+\V j^{\rm mol}_\xi)\over \partial {\V r}}  ={I}^{\rm gain}.
\label{conserv}
\ee 
consisting of a quasiparticle part 
\be
\xi^{\rm qp}=\int{d^3k\over (2\pi)^3} \xi_1 f_1 
\ee 
and the correlated or molecular contribution
\ba
\xi^{\rm mol}&=\int{d^3 \! kd^3 \! pd^3 \! q\over(2\pi)^9}
D\Delta_t {\xi_1+\xi_2\over 2}.
\label{ob25}
\end{align}
The latter one leads to the statistical interpretation as if two particles form a molecule. The rate of binary processes $D=\left|{\cal T}\right|^22\pi\delta(\varepsilon_1\!+\!\varepsilon_2\!-\!
\varepsilon_3\!-\!\varepsilon_4\!+\!2\Delta_E)\left(1\!-\!f_3\!-\!f_4\right)f_1f_2$ is weighed with the lifetime of the molecule $\Delta_t$, respectively. 
We suppress the obvious sum over sort indices for the sake of legibility.

The usual quasiparticle currents of the observables read
\begin{equation}
\V {j}_\xi^{\mathrm{qp}}=\int{d^3 {k}\over(2\pi)^3}\xi_1
{\partial\varepsilon_1\over\partial \V {k}}f_1
\label{ob51a}
\end{equation}
and the molecular currents we have obtained as \cite{LSM97,M17,M17b}
\ba
\V j_\xi^{\rm mol}=\frac 1 2 \!\int\!\!{d^3 \! kd^3 \! pd^3 \! q\over(2\pi)^9}D(\xi_2 {\V \Delta_2}\!-\!\xi_3 {\V \Delta_3}\!-\!\xi_4 {\V \Delta_4}).
\label{ecurrmol}
\end{align}
It is the balance of observables carried by the different spatial off-sets.

To present explicit formulas, the total quasiparticle stress tensor formed by the quasiparticles read 
\begin{equation}
{\Pi}_{ij}^{\rm qp}=\int{d^3 \! k\over(2\pi)^3}\left(k_j
{\partial\varepsilon\over\partial k_i}+\delta_{ij}\varepsilon\right)f-
\delta_{ij}{E}^{\rm qp}
\label{ec81}
\end{equation}
with the quasiparticle energy functional \cite{LSM97}
\ba
{E}^{\rm qp}&=\int{d^3 \! k\over(2\pi)^3}f_1(k){k^2\over 2m}
\nonumber\\+&
{1\over 2}\int{d^3 \! kd^3 \! p\over(2\pi)^6}f_1(k)f_2(p)
{\rm Re}\,  {\cal T}(\varepsilon_1+\varepsilon_2,k,p,0)
\label{ec22cb}
\end{align}
instead of the Landau functional
which is valid only in local approximation.

The molecular contributions (\ref{ecurrmol}) to the stress tensor reads 
\ba
{\Pi}_{ij}^{\rm mol}=&{1\over 2}\int{d^3 \! kd^3 \! pd^3 \! q\over(2\pi)^9}
D
\nonumber\\&\times
\left [(k_j-q_j) \Delta_{3i}+(p_j+q_j )\Delta_{4i}-p_j \Delta_{2i}\right ].
\label{ec82}
\end{align}
This momentum tensor is the balance of the momenta carried by the corresponding spatial off-sets weighted with the rate to form a molecule $D$.

For the density $\xi=1$ we do not have a gain $I^{\rm gain}$. The momentum gain
$\xi=k_j$ reads 
\begin{equation}
{I}^{\rm gain}_{Kj}=\int{d^3 \! kd^3 \! pd^3 \! q\over(2\pi)^9}
D
\Delta_{Kj}
.
\label{gainK}
\end{equation}
Dividing and multiplying by $\Delta_t$ under the integral we see that the
momentum gain is the probability $D\Delta_t$ to form a molecule multiplied with the force $\V \Delta_K/\Delta_t$ exercised during the delay time $\Delta_t$ from the environment by all other particles.
This momentum gain
(\ref{gainK}) can be exactly recast together with the term of the drift into a spatial derivative \cite{LSM97,M17b}
\be
\int{d^3 \! k\over(2\pi)^3}\varepsilon{\partial f\over\partial \V r_j}
+{I}^{\rm gain}_{Kj}={\partial{E}^{\rm qp}\over\partial \V r_j}
\label{ec79}
\ee
of the quasiparticle energy functional (\ref{ec22cb}).
Analogously the energy gain 
\be
{I}^{\rm gain}_E=\int{d^3 \! kd^3 \! pd^3 \! q\over(2\pi)^9}
D
\Delta_{E}
\label{gainE}
\ee
combines with the drift into the total time derivative of the quasiparticle energy functional (\ref{ec22cb})
\begin{equation}
\int{d^3 \! k\over(2\pi)^3}\varepsilon
{\partial f\over\partial t}
-{I}^{\rm gain}_E=
{\partial{E}^{\rm qp}\over\partial t}.
\label{ec11}
\end{equation}

The only remaining explicit gain is the entropy gain
\ba
&{I}^{\rm gain}_S
=
-\frac{k_B}{2} \int{d^3 \! kd^3 \! pd^3 \! q\over(2\pi)^9}
D \ln {{f_3 f_4 (1\!-\!f_1)(1\!-\!f_2)\over (1\!-\!f_3)(1\!-\!f_4) f_1 f_2}}
\label{gainentrop}
\end{align}
while the momentum gain and energy gain are transferring kinetic into correlation parts and do not appear explicitly.
In \cite{M17} it is proved that this entropy gain is always positive establishing the H-theorem including single particle and two-particle quantum correlations.
In other words, the additional gain on the right side of the balance equations (\ref{conserv}) might be due
to an energy or force feed from the outside or the entropy production by collisions.

\section{Model with time-dependent interaction}\label{LowenergyT}
So far we have summarized the nonlocal extensions of the quantum Boltzmann equation. Now, as exploratory example and new result, we want to consider a
point-like interaction where the T-matrix is dominated by the s-wave channel with a single
time-dependent scattering length $a_\mathrm{sc}(t)$.
One might think on cold atoms where the time-dependent magnetic field $\V B(\V r,t)$ near the Feshbach resonance determines this scattering length \cite{Chin10}
\begin{equation}\label{Feshbach}
a_\mathrm{sc}(\V r,t)\approx a_\mathrm{bg}\left(1-\frac{\Gamma_i}{\left|\V B(\V r,t)\right|-B_i}\right).
\end{equation}
In such a way the binary T-matrix becomes an externally controlled function of time and space.
We consider two atoms with their center-of-the-mass momentum $K=k+p$ and their difference momentum $\kappa={m_b\over M} k-{m_a\over M} p$ where $M=m_a+m_b$ and $\mu^{-1}=m_a^{-1}+m_b^{-1}$. 
The sum of energies before $(k,p)$ and after the collision $(k-q,p+q)$ reads
\be 
\epsilon_1+\epsilon_2&=&{K^2\over 2 M}+{\kappa^2\over 2\mu} 
\nonumber\\
\epsilon_3+\epsilon_4&=&{K^2\over 2 M}+{\kappa^2+q^2\over 2\mu}-{\V q\cdot \V \kappa\over \mu} 
\ee

In the dilute gas the medium effect on the binary
interaction caused by the Pauli blocking are negligible so that we can
use the free-space T-matrix
\begin{equation}\label{Tfree}
{\cal T}_R(t)=\frac{2\pi \hbar^2 a_\mathrm{sc}(t)}{\mu}\frac {1}{1+i{a_\mathrm{sc}(t)\over \hbar}\sqrt{2\mu(\Omega-K^2/2M)}}
\end{equation}
which one obtains by solving the T-matrix equation with running coupling constant \cite{M09}. 
The corresponding set of $\Delta$s given by Eqs.~\eqref{vc4g} becomes due to the point-like interaction
\begin{eqnarray}
\Delta_t(t)&=&-\frac{a_\mathrm{sc}\mu}{\kappa \left (1+{a^2_\mathrm{sc}\kappa^2/\hbar^2}\right )},
\nonumber\\
{\V \Delta_K}(t)&=&{1\over a_\mathrm{sc}}{\partial a_\mathrm{sc}\over\partial \V r}{\kappa^2\over 2 \mu}\Delta_t,\qquad
\Delta_E(t)=-{1\over a_\mathrm{sc}}\frac{\partial a_\mathrm{sc}}{\partial t}{\kappa^2\over 2 \mu}\Delta_t
\nonumber\\
{\V \Delta_3}(t)&=&{\V \Delta_4}=\frac{\V K}{M}\Delta_t,\qquad
\V \Delta_2=0,
\label{vc4gfree}
\end{eqnarray}
The value of $\V\Delta_{3,4}$ is a free flight 
of the interacting pair during $\Delta_t$.

Now we calculate all thermodynamic quantities.
Since the T-matrix and all shifts are independent of the transferred momentum $q$ one can integrate easily 
\be
\int {d^3 q\over (2\pi\hbar)^3} \delta (\epsilon_1+\epsilon_2-\epsilon_3-\epsilon_4)={\mu \kappa\over 2 \pi^2\hbar^3}.
\ee
For Maxwellian distributions of equal temperature $T$ and densities of the two species $n_a$ and $n_b$,
\be
f_1f_2=n_a n_b {(2\pi\hbar^2)^3\over T^3 (\mu M)^{3/2}}{\rm e}^{-{K^2\over 2 M T}-{\kappa^2\over 2 \mu T}},
\ee
one can perform trivially the $K$ integration since all shifts and the T-matrix are only dependent on $\kappa$.
Then the correlated density, energy (\ref{ob25}), and momentum tensor (\ref{ec82}) become
\be
n^{\rm mol}(t)&=&\frac 1 2 \int {d\kappa\over (2\pi\hbar)^3} \tilde D,\quad \Pi_{ij}^{\rm mol}(t)=\delta_{ij} T n^{\rm mol},\nonumber\\
 E^{\rm mol}&=&\frac 1 2 \int {d\kappa\over (2\pi\hbar)^3} \tilde D\left (\frac 3 2 T+{\kappa^2\over 2\mu}\right )
\ee
and the energy gain (\ref{gainE}) and momentum gain (\ref{gainK}) are
\be
\left .\begin{matrix}
I_E^{\rm gain}(t)
\cr
\V I_K^{\rm gain}(t)
\end{matrix}\right \}= \int {d\kappa\over (2\pi\hbar)^3} \tilde D {\kappa^2\over 2 \mu}
\left \{\begin{matrix}-\frac{\partial a_\mathrm{sc}}{\partial t}
\cr
\frac{\partial a_\mathrm{sc}}{\partial \V r}\end{matrix}\right .
\ee
where we used
\be
\tilde D=-{4\pi^{5/2}\hbar^6\over \mu^3T^3}n_an_b x^3{{\rm e}^{-{\kappa^2\over 2\mu T}}\over  \left (1+{a_\mathrm{sc}}^2\kappa^2/ \hbar^2\right )^2}
\ee
and introduce with the De Broglie wavelength $\lambda^2=2\pi \hbar^2/2 \mu T$ the time-dependent variable
\be
x^2(t)=2\pi {a_\mathrm{sc}^2(t)\over \lambda^2}.
\label{x}
\ee
All correlated currents are zero in this example.

The various $\kappa$ integrations are straightforward with the help of the error function
\be
x \xi(x)=\sqrt{2\over \pi}\int\limits_0^\infty d t {{\rm e}^{-{t^2\over x^2}}\over 1+t^2}=\sqrt{\pi \over 2}{\rm e}^{x^{-2}}{\rm erfc}\left (x^{-1} \right ).
\ee

The final results read then
\be
n^{\rm mol}(x)&=& n_a n_b {\pi\hbar^3\over (\mu T)^{3/2}} x^2 \xi'(x),
\nonumber\\ 
\Pi^{\rm mol}_{ij}(x)&=&\delta _{ij} T n^{mol}(x)
\nonumber\\
E^{\rm mol}(x)&=&\frac 3 2 n^{\rm mol}(x) T+T x^2\left ({n^{\rm mol}(x)\over x}\right )'
\nonumber\\
\left .\begin{matrix}
I_E^{\rm gain}(x)
\cr
\V I_K^{\rm gain}(x)
\end{matrix}\right \}&=& 2 T x^2\left ({n^{\rm mol}(x)\over x}\right )'
\left \{\begin{matrix}-\frac{\partial a_\mathrm{sc}}{\partial t}
\cr
\frac{\partial a_\mathrm{sc}}{\partial \V r}\end{matrix}\right .
\label{final}
\ee
where we denote the prime as derivative with respect to $x$.
The quasiparticle energy (\ref{ec22cb})
becomes
\be
E^{\rm qp}(x)=\frac 3 2 T (n_a+n_b)+4 T n^{\rm mol}(x)
\label{eqp}
\ee
which shows the expected three translational degrees of freedom for the free particles and 8 degrees of freedom for the correlated molecules. The latter can be understood as twice the three translational degrees of freedom of the two colliding particles and 2 additional rotational degrees of freedom if the two particles form a bound state seen as classical dumbbell.

The molecular energy behaves as
\begin{eqnarray}
E^\mathrm{mol}
&=&n^{\rm mol} T 
\left \{
\begin{matrix}
{1 \over 2}+{4\over \sqrt{\pi} x}+o(x^{-2})
\cr
{7\over 2} -6 x^2+o(x^3)
\end{matrix}
\right .
\label{high}
\end{eqnarray}
in the high and low temperature limit, respectively (\ref{x}).
Compared  with the quasiparticle part (\ref{eqp}) the high-temperature limit (\ref{high}) brings an additional degree of freedom by the correlational contribution. It can be seen as an additional internal degree of freedom like the two-particle dumbbell state gains an additional third fictitious particle by correlations.

\section{Effect of external power feed}
In order to check the energy conservation (\ref{ec11}) we consider the effect of the external power feed due to the time-dependent potential ${V}(t)=2\pi\hbar^2 a_\mathrm{sc}(t) /\mu$. From the retarded and advanced T-matrices in operator notation ${\cal T}^{-1}_{R/A}=V(t)^{-1}-G_{R/A}$ follows that one can write the real part of the T-matrix ${\rm Re} {\cal T}=({\cal T}_R+{\cal T}_A)/2={\cal T}_R(V^{-1}-{\rm Re} G){\cal T}_A$. From (\ref{ec22cb}) we see therefore that the time-dependent potential leads to an extra feed 
\be
w_E&=&
-\frac 1 2 \int{d^3 k d^3p\over (2\pi\hbar)^3} {\partial {V}^{-1}\over \partial t} |{\cal T}|^2 f_1 f_2
\label{pumpE}
\ee
to the energy balance (\ref{gainE})  
\be
{\partial E^{\rm qp}\over \partial t}=-I_E^{\rm gain}+w_E
\label{last}
\ee
since 
$\epsilon=k^2/2m_a$ and $f$ are independent of time here. For the considered point-interaction model we get
\be
w_E
&=&4 T n^{\rm mol} {\partial \ln a_\mathrm{sc}\over \partial t}.
\ee
On the other hand, the derivative of (\ref{eqp}) can be calculated explicitly
\ba
{\partial E^{\rm qp}\over \partial t} =4 T x (n^{\rm mol})' \frac {1}{a_\mathrm{sc}}{\partial a_\mathrm{sc}\over \partial t}=-I_E^{\rm gain}+4 T n^{\rm mol}{\partial \ln a_\mathrm{sc}\over \partial t} 
\label{proof1}
\end{align}
and one sees how the extra feed  (\ref{pumpE}) appears such that indeed (\ref{last}) holds. This illustrates the proof of energy conservation (\ref{ec11}). 

\begin{figure}[]
  \includegraphics[width=9.5cm]{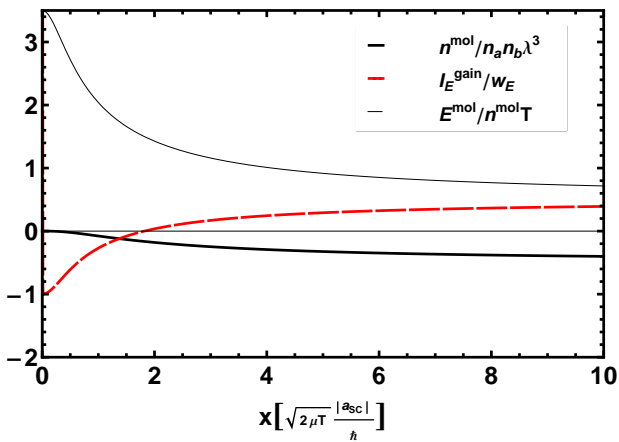}
\caption{The molecular density (\ref{ob25}) (black thick line), energy gain (\ref{gainE}) per external power (\ref{pumpE}) which is the latent heat (red broken line) and correlation energy (\ref{ob25}) (thin line) versus the dimensionless scattering length (\ref{x}) for Maxwellian particles.
}
\label{corrvar}
\end{figure}

The correlated density in (\ref{final}) appears as the Beth-Uhlenbeck equation of state \cite{BU37,SRS90}.
The correlated density is negative according to the fact that we can only describe attractive interactions with contact potentials \cite{M09}. Then the two correlated particles form possible bound states.
The density of correlated particles or molecules possesses 
the low-temperature limit
\be
n^{\rm mol}\!=\!-\! n_a n_b \lambda^3 {\sqrt{\pi}\over 4} x^3\!+\!o(x^4)=\!-\!2\pi^2 n_a n_b a_\mathrm{sc}^3\!+\!o(T^2)
\ee 
which is just the expression of the second virial coefficient for hard spheres to the pressure.
The high-temperature limit, $n^{\rm mol}(x)
= - n_a n_b \lambda^3{\pi \over 4}+o(x^{-1})$,
vanishes with $T^{-3/2}$.

In figure~\ref{corrvar} we plot the ratio of the energy gain (\ref{gainE}) to the external power (\ref{pumpE}) with the analytical results (\ref{final}). This has the merit to be a dimensionless quantity where the time derivative $\partial_t \ln a_\mathrm{sc}(t)$ drops out. It is now easy to see that the energy gain (\ref{gainE}) and entropy gain (\ref{gainentrop}) are linked for equilibrium distributions as
\be
T I_S^{\rm gain}={I_E^{\rm gain}}.
\ee
Since the latent heat is the temperature times the entropy difference occurring during a phase transition we can consider the formation of short-living molecules here analogously. Therefore the energy gain is the rate of latent heat. Dividing by the external power we obtain the ratio of the latent heat to the interaction strength due to correlations.

It is remarkable that the energy gain relative to the external pumping changes the sign at $x_0\approx 1.8184$ which means $a_\mathrm{sc}/\lambda\approx 0.7254$. This value is independent of interaction and in this sense universal for such short-range interactions. For products of scattering length and temperatures smaller than this value the correlations lead to a behavior oppositely as expected from the feed, $I_E^{\rm gain}/w_E=-1+3 x^2 +o(x)$ and we have correlational cooling. For high temperatures the gain approaches half of the external power $I_E^{\rm gain}/w_E=1/2-2/\sqrt{\pi}x+o(x^{-2})$ and we see correlational heating.

\section{Summary}
To summarize, when strong correlations are formed there is a cancellation of off-shell processes in the kinetic equation resulting into a proper extended quasiparticle picture. The remaining modifications of the quantum Boltzmann equation consist in the nonlocal collision scenario where the off-sets are uniquely determined by the phase shift of the T-matrix and the quasiparticle energies modifying the drift. The resulting balance equations show besides the quasiparticle parts of the Landau theory also explicit two-particle contributions of short living molecules. The energy and momentum conservation is ensured due to an internal transfer of energy and momentum analogously to a latent heat. Only for the entropy an explicit gain remains which can be proved to be positive \cite{M17} ensuring Boltzmann's H-theorem. The single-particle entropy can decrease on cost of the correlated part of entropy describing the two-particles in a molecular state. 

For an exploratory example of Maxwellian particles we find a sign change of the energy gain compared with the external power feed independent of the interaction. The interpretation as the rate of latent heat due to correlations is suggested leading to correlational heating and cooling.

\bibliography{entropy,bose,kmsr,kmsr1,kmsr2,kmsr3,kmsr4,kmsr5,kmsr6,kmsr7,delay2,spin,spin1,refer,delay3,gdr,chaos,sem3,sem1,sem2,short,cauchy,genn,paradox,deform,shuttling,blase,spinhall,spincurrent,tdgl,pattern,zitter,graphene,quench,msc_nodouble,iso,march}

\begin{thebibliography}{48}
\expandafter\ifx\csname natexlab\endcsname\relax\def\natexlab#1{#1}\fi
\expandafter\ifx\csname bibnamefont\endcsname\relax
  \def\bibnamefont#1{#1}\fi
\expandafter\ifx\csname bibfnamefont\endcsname\relax
  \def\bibfnamefont#1{#1}\fi
\expandafter\ifx\csname citenamefont\endcsname\relax
  \def\citenamefont#1{#1}\fi
\expandafter\ifx\csname url\endcsname\relax
  \def\url#1{\texttt{#1}}\fi
\expandafter\ifx\csname urlprefix\endcsname\relax\def\urlprefix{URL }\fi
\providecommand{\bibinfo}[2]{#2}
\providecommand{\eprint}[2][]{\url{#2}}

\bibitem[{\citenamefont{Chapman and Cowling}(1990)}]{CC90}
\bibinfo{author}{\bibfnamefont{S.}~\bibnamefont{Chapman}} \bibnamefont{and}
  \bibinfo{author}{\bibfnamefont{T.~G.} \bibnamefont{Cowling}},
  \emph{\bibinfo{title}{The Mathematical Theory of Non-uniform Gases}}
  (\bibinfo{publisher}{Cambrigde University Press},
  \bibinfo{address}{Cambridge}, \bibinfo{year}{1990}), \bibinfo{note}{third
  edition Chap. 16}.

\bibitem[{\citenamefont{Alexander et~al.}(1995)\citenamefont{Alexander, Garcia,
  and Alder}}]{AGA95}
\bibinfo{author}{\bibfnamefont{F.~J.} \bibnamefont{Alexander}},
  \bibinfo{author}{\bibfnamefont{A.~L.} \bibnamefont{Garcia}},
  \bibnamefont{and} \bibinfo{author}{\bibfnamefont{B.~J.} \bibnamefont{Alder}},
  \bibinfo{journal}{Phys. Rev. Lett.} \textbf{\bibinfo{volume}{74}},
  \bibinfo{pages}{5212} (\bibinfo{year}{1995}).

\bibitem[{\citenamefont{Waldmann}(1960)}]{W60}
\bibinfo{author}{\bibfnamefont{L.}~\bibnamefont{Waldmann}},
  \bibinfo{journal}{Z. Naturforsch. a} \textbf{\bibinfo{volume}{15}},
  \bibinfo{pages}{19} (\bibinfo{year}{1960}).

\bibitem[{\citenamefont{Snider}(1964)}]{S64}
\bibinfo{author}{\bibfnamefont{R.~F.} \bibnamefont{Snider}},
  \bibinfo{journal}{J. Math. Phys.} \textbf{\bibinfo{volume}{5}},
  \bibinfo{pages}{1580} (\bibinfo{year}{1964}).

\bibitem[{\citenamefont{B{\"a}rwinkel}(1969)}]{B69}
\bibinfo{author}{\bibfnamefont{K.}~\bibnamefont{B{\"a}rwinkel}},
  \bibinfo{journal}{Z. Naturforsch.} \textbf{\bibinfo{volume}{24a}},
  \bibinfo{pages}{38} (\bibinfo{year}{1969}).

\bibitem[{\citenamefont{Thomas and Snider}(1970)}]{TS70}
\bibinfo{author}{\bibfnamefont{M.~W.} \bibnamefont{Thomas}} \bibnamefont{and}
  \bibinfo{author}{\bibfnamefont{R.~F.} \bibnamefont{Snider}},
  \bibinfo{journal}{J. Stat. Phys.} \textbf{\bibinfo{volume}{2}},
  \bibinfo{pages}{61} (\bibinfo{year}{1970}).

\bibitem[{\citenamefont{Snider and Sanctuary}(1971)}]{SS71}
\bibinfo{author}{\bibfnamefont{R.~F.} \bibnamefont{Snider}} \bibnamefont{and}
  \bibinfo{author}{\bibfnamefont{B.~C.} \bibnamefont{Sanctuary}},
  \bibinfo{journal}{J. Chem. Phys.} \textbf{\bibinfo{volume}{55}},
  \bibinfo{pages}{1555} (\bibinfo{year}{1971}).

\bibitem[{\citenamefont{Rainwater and Snider}(1976)}]{RS76}
\bibinfo{author}{\bibfnamefont{J.~C.} \bibnamefont{Rainwater}}
  \bibnamefont{and} \bibinfo{author}{\bibfnamefont{R.~F.}
  \bibnamefont{Snider}}, \bibinfo{journal}{J. Chem. Phys.}
  \textbf{\bibinfo{volume}{65}}, \bibinfo{pages}{4958} (\bibinfo{year}{1976}).

\bibitem[{\citenamefont{Balescu}(1975)}]{B75}
\bibinfo{author}{\bibfnamefont{R.}~\bibnamefont{Balescu}},
  \emph{\bibinfo{title}{Equilibrium and Nonequilibrium Statistically
  Mechanics}} (\bibinfo{publisher}{Wiley}, \bibinfo{address}{New York},
  \bibinfo{year}{1975}).

\bibitem[{\citenamefont{McLennan}(1989)}]{M89}
\bibinfo{author}{\bibfnamefont{J.~A.} \bibnamefont{McLennan}},
  \emph{\bibinfo{title}{Introduction to Nonequilibrium Statistical Mechanics}}
  (\bibinfo{publisher}{Prentice-Hall}, \bibinfo{address}{Englewood Cliffs},
  \bibinfo{year}{1989}).

\bibitem[{\citenamefont{Lalo{\"e}}(1989)}]{La89}
\bibinfo{author}{\bibfnamefont{F.}~\bibnamefont{Lalo{\"e}}},
  \bibinfo{journal}{J. Phys. (Paris)} \textbf{\bibinfo{volume}{50}},
  \bibinfo{pages}{1851} (\bibinfo{year}{1989}).

\bibitem[{\citenamefont{Nacher et~al.}(1989)\citenamefont{Nacher, Tastevin, and
  Lalo{\"e}}}]{NTL89}
\bibinfo{author}{\bibfnamefont{P.}~\bibnamefont{Nacher}},
  \bibinfo{author}{\bibfnamefont{G.}~\bibnamefont{Tastevin}}, \bibnamefont{and}
  \bibinfo{author}{\bibfnamefont{F.}~\bibnamefont{Lalo{\"e}}},
  \bibinfo{journal}{J. Phys. (Paris)} \textbf{\bibinfo{volume}{50}},
  \bibinfo{pages}{1907} (\bibinfo{year}{1989}).

\bibitem[{\citenamefont{Loos}(1990)}]{L90a}
\bibinfo{author}{\bibfnamefont{D.}~\bibnamefont{Loos}}, \bibinfo{journal}{J.
  Stat. Phys.} \textbf{\bibinfo{volume}{61}}, \bibinfo{pages}{467}
  (\bibinfo{year}{1990}).

\bibitem[{\citenamefont{de~Haan}(1991)}]{H91}
\bibinfo{author}{\bibfnamefont{H.}~\bibnamefont{de~Haan}},
  \bibinfo{journal}{Physica A} \textbf{\bibinfo{volume}{170}},
  \bibinfo{pages}{571} (\bibinfo{year}{1991}).

\bibitem[{\citenamefont{Lalo{\"e} and Mullin}(1990)}]{LM90}
\bibinfo{author}{\bibfnamefont{F.}~\bibnamefont{Lalo{\"e}}} \bibnamefont{and}
  \bibinfo{author}{\bibfnamefont{W.~J.} \bibnamefont{Mullin}},
  \bibinfo{journal}{J. Stat. Phys.} \textbf{\bibinfo{volume}{59}},
  \bibinfo{pages}{725} (\bibinfo{year}{1990}).

\bibitem[{\citenamefont{Nacher et~al.}(1991{\natexlab{a}})\citenamefont{Nacher,
  Tastevin, and Lalo{\"e}}}]{NTL91}
\bibinfo{author}{\bibfnamefont{P.~J.} \bibnamefont{Nacher}},
  \bibinfo{author}{\bibfnamefont{G.}~\bibnamefont{Tastevin}}, \bibnamefont{and}
  \bibinfo{author}{\bibfnamefont{F.}~\bibnamefont{Lalo{\"e}}},
  \bibinfo{journal}{Ann. Phys. (Leipzig)} \textbf{\bibinfo{volume}{48}},
  \bibinfo{pages}{149} (\bibinfo{year}{1991}{\natexlab{a}}).

\bibitem[{\citenamefont{Nacher et~al.}(1991{\natexlab{b}})\citenamefont{Nacher,
  Tastevin, and Lalo{\"e}}}]{NTL91a}
\bibinfo{author}{\bibfnamefont{P.~J.} \bibnamefont{Nacher}},
  \bibinfo{author}{\bibfnamefont{G.}~\bibnamefont{Tastevin}}, \bibnamefont{and}
  \bibinfo{author}{\bibfnamefont{F.}~\bibnamefont{Lalo{\"e}}},
  \bibinfo{journal}{Journal de Physique I} \textbf{\bibinfo{volume}{1}},
  \bibinfo{pages}{181} (\bibinfo{year}{1991}{\natexlab{b}}).

\bibitem[{\citenamefont{Snider}(1995)}]{S95}
\bibinfo{author}{\bibfnamefont{R.~F.} \bibnamefont{Snider}},
  \bibinfo{journal}{J. Stat. Phys.} \textbf{\bibinfo{volume}{80}},
  \bibinfo{pages}{1085} (\bibinfo{year}{1995}).

\bibitem[{\citenamefont{Snider et~al.}(1995)\citenamefont{Snider, Mullin, and
  Lalo{\"e}}}]{SML95}
\bibinfo{author}{\bibfnamefont{R.~F.} \bibnamefont{Snider}},
  \bibinfo{author}{\bibfnamefont{W.~J.} \bibnamefont{Mullin}},
  \bibnamefont{and}
  \bibinfo{author}{\bibfnamefont{F.}~\bibnamefont{Lalo{\"e}}},
  \bibinfo{journal}{Physica A} \textbf{\bibinfo{volume}{218}},
  \bibinfo{pages}{155} (\bibinfo{year}{1995}).

\bibitem[{\citenamefont{{\v S}pi{\v c}ka et~al.}(1998)\citenamefont{{\v S}pi{\v
  c}ka, Lipavsk{\'y}, and Morawetz}}]{SLM96}
\bibinfo{author}{\bibfnamefont{V.}~\bibnamefont{{\v S}pi{\v c}ka}},
  \bibinfo{author}{\bibfnamefont{P.}~\bibnamefont{Lipavsk{\'y}}},
  \bibnamefont{and} \bibinfo{author}{\bibfnamefont{K.}~\bibnamefont{Morawetz}},
  \bibinfo{journal}{Phys. Lett. A} \textbf{\bibinfo{volume}{240}},
  \bibinfo{pages}{160} (\bibinfo{year}{1998}).

\bibitem[{\citenamefont{Morawetz
  et~al.}(2001{\natexlab{a}})\citenamefont{Morawetz, Lipavsk{\'y}, and {\v
  S}pi{\v c}ka}}]{MLS00}
\bibinfo{author}{\bibfnamefont{K.}~\bibnamefont{Morawetz}},
  \bibinfo{author}{\bibfnamefont{P.}~\bibnamefont{Lipavsk{\'y}}},
  \bibnamefont{and} \bibinfo{author}{\bibfnamefont{V.}~\bibnamefont{{\v S}pi{\v
  c}ka}}, \bibinfo{journal}{Ann. of Phys.} \textbf{\bibinfo{volume}{294}},
  \bibinfo{pages}{134} (\bibinfo{year}{2001}{\natexlab{a}}).

\bibitem[{\citenamefont{Lipavsk{\'y} et~al.}(2001)\citenamefont{Lipavsk{\'y},
  Morawetz, and {\v S}pi{\v c}ka}}]{LSM97}
\bibinfo{author}{\bibfnamefont{P.}~\bibnamefont{Lipavsk{\'y}}},
  \bibinfo{author}{\bibfnamefont{K.}~\bibnamefont{Morawetz}}, \bibnamefont{and}
  \bibinfo{author}{\bibfnamefont{V.}~\bibnamefont{{\v S}pi{\v c}ka}},
  \emph{\bibinfo{title}{Kinetic equation for strongly interacting dense Fermi
  systems}}, vol. \bibinfo{volume}{26,1} of \emph{\bibinfo{series}{Annales de
  Physique}} (\bibinfo{publisher}{EDP Sciences}, \bibinfo{address}{Paris},
  \bibinfo{year}{2001}).

\bibitem[{\citenamefont{Craig}(1966)}]{C66a}
\bibinfo{author}{\bibfnamefont{R.~A.} \bibnamefont{Craig}},
  \bibinfo{journal}{Ann. Phys.} \textbf{\bibinfo{volume}{40}},
  \bibinfo{pages}{416} (\bibinfo{year}{1966}).

\bibitem[{\citenamefont{Bezzerides and DuBois}(1968)}]{BD68}
\bibinfo{author}{\bibfnamefont{B.}~\bibnamefont{Bezzerides}} \bibnamefont{and}
  \bibinfo{author}{\bibfnamefont{D.~F.} \bibnamefont{DuBois}},
  \bibinfo{journal}{Phys. Rev.} \textbf{\bibinfo{volume}{168}},
  \bibinfo{pages}{233} (\bibinfo{year}{1968}).

\bibitem[{\citenamefont{Stolz and Zimmermann}(1979)}]{SZ79}
\bibinfo{author}{\bibfnamefont{H.}~\bibnamefont{Stolz}} \bibnamefont{and}
  \bibinfo{author}{\bibfnamefont{R.}~\bibnamefont{Zimmermann}},
  \bibinfo{journal}{phys. stat. sol. (b)} \textbf{\bibinfo{volume}{94}},
  \bibinfo{pages}{135} (\bibinfo{year}{1979}).

\bibitem[{\citenamefont{Kremp et~al.}(1984)\citenamefont{Kremp, Kraeft, and
  Lambert}}]{KKL84}
\bibinfo{author}{\bibfnamefont{D.}~\bibnamefont{Kremp}},
  \bibinfo{author}{\bibfnamefont{W.~D.} \bibnamefont{Kraeft}},
  \bibnamefont{and} \bibinfo{author}{\bibfnamefont{A.~D.~J.}
  \bibnamefont{Lambert}}, \bibinfo{journal}{Physica A}
  \textbf{\bibinfo{volume}{127}}, \bibinfo{pages}{72} (\bibinfo{year}{1984}).

\bibitem[{\citenamefont{Schmidt and R{\"o}pke}(1987)}]{SR87}
\bibinfo{author}{\bibfnamefont{M.}~\bibnamefont{Schmidt}} \bibnamefont{and}
  \bibinfo{author}{\bibfnamefont{G.}~\bibnamefont{R{\"o}pke}},
  \bibinfo{journal}{phys. stat. sol. (b)} \textbf{\bibinfo{volume}{139}},
  \bibinfo{pages}{441} (\bibinfo{year}{1987}).

\bibitem[{\citenamefont{K{\"o}hler and Malfliet}(1993)}]{KM93}
\bibinfo{author}{\bibfnamefont{H.~S.} \bibnamefont{K{\"o}hler}}
  \bibnamefont{and} \bibinfo{author}{\bibfnamefont{R.}~\bibnamefont{Malfliet}},
  \bibinfo{journal}{Phys. Rev. C} \textbf{\bibinfo{volume}{48}},
  \bibinfo{pages}{1034} (\bibinfo{year}{1993}).

\bibitem[{\citenamefont{{\v S}pi{\v c}ka and Lipavsk{\'y}}(1995)}]{SL95}
\bibinfo{author}{\bibfnamefont{V.}~\bibnamefont{{\v S}pi{\v c}ka}}
  \bibnamefont{and}
  \bibinfo{author}{\bibfnamefont{P.}~\bibnamefont{Lipavsk{\'y}}},
  \bibinfo{journal}{Phys. Rev. B} \textbf{\bibinfo{volume}{52}},
  \bibinfo{pages}{14615} (\bibinfo{year}{1995}).

\bibitem[{\citenamefont{{\v S}pi{\v c}ka et~al.}(1997)\citenamefont{{\v S}pi{\v
  c}ka, Lipavsk{\'y}, and Morawetz}}]{SLMa96}
\bibinfo{author}{\bibfnamefont{V.}~\bibnamefont{{\v S}pi{\v c}ka}},
  \bibinfo{author}{\bibfnamefont{P.}~\bibnamefont{Lipavsk{\'y}}},
  \bibnamefont{and} \bibinfo{author}{\bibfnamefont{K.}~\bibnamefont{Morawetz}},
  \bibinfo{journal}{Phys. Rev. B} \textbf{\bibinfo{volume}{55}},
  \bibinfo{pages}{5084} (\bibinfo{year}{1997}).

\bibitem[{\citenamefont{K{\"o}hler}(1995)}]{K95}
\bibinfo{author}{\bibfnamefont{H.~S.} \bibnamefont{K{\"o}hler}},
  \bibinfo{journal}{Phys. Rev. C} \textbf{\bibinfo{volume}{51}},
  \bibinfo{pages}{3232} (\bibinfo{year}{1995}).

\bibitem[{\citenamefont{Morawetz
  et~al.}(1999{\natexlab{a}})\citenamefont{Morawetz, Lipavsk{\'y}, {\v S}pi{\v
  c}ka, and Kwong}}]{MLSK99}
\bibinfo{author}{\bibfnamefont{K.}~\bibnamefont{Morawetz}},
  \bibinfo{author}{\bibfnamefont{P.}~\bibnamefont{Lipavsk{\'y}}},
  \bibinfo{author}{\bibfnamefont{V.}~\bibnamefont{{\v S}pi{\v c}ka}},
  \bibnamefont{and} \bibinfo{author}{\bibfnamefont{N.-H.} \bibnamefont{Kwong}},
  \bibinfo{journal}{Phys. Rev. C} \textbf{\bibinfo{volume}{59}},
  \bibinfo{pages}{3052} (\bibinfo{year}{1999}{\natexlab{a}}).

\bibitem[{\citenamefont{Schmidt et~al.}(1990)\citenamefont{Schmidt, R{\"o}pke,
  and Schulz}}]{SRS90}
\bibinfo{author}{\bibfnamefont{M.}~\bibnamefont{Schmidt}},
  \bibinfo{author}{\bibfnamefont{G.}~\bibnamefont{R{\"o}pke}},
  \bibnamefont{and} \bibinfo{author}{\bibfnamefont{H.}~\bibnamefont{Schulz}},
  \bibinfo{journal}{Ann. Phys. (NY)} \textbf{\bibinfo{volume}{202}},
  \bibinfo{pages}{57} (\bibinfo{year}{1990}).

\bibitem[{\citenamefont{Morawetz and Roepke}(1995)}]{MR94}
\bibinfo{author}{\bibfnamefont{K.}~\bibnamefont{Morawetz}} \bibnamefont{and}
  \bibinfo{author}{\bibfnamefont{G.}~\bibnamefont{Roepke}},
  \bibinfo{journal}{Phys. Rev. E} \textbf{\bibinfo{volume}{51}},
  \bibinfo{pages}{4246} (\bibinfo{year}{1995}).

\bibitem[{\citenamefont{Lipavsk{\'y} et~al.}(1999)\citenamefont{Lipavsk{\'y},
  {\v S}pi{\v c}ka, and Morawetz}}]{LSM99}
\bibinfo{author}{\bibfnamefont{P.}~\bibnamefont{Lipavsk{\'y}}},
  \bibinfo{author}{\bibfnamefont{V.}~\bibnamefont{{\v S}pi{\v c}ka}},
  \bibnamefont{and} \bibinfo{author}{\bibfnamefont{K.}~\bibnamefont{Morawetz}},
  \bibinfo{journal}{Phys. Rev. E} \textbf{\bibinfo{volume}{59}},
  \bibinfo{pages}{{R1291}} (\bibinfo{year}{1999}).

\bibitem[{\citenamefont{{\v S}pi{\v c}ka et~al.}(2001)\citenamefont{{\v S}pi{\v
  c}ka, Morawetz, and Lipavsk{\'y}}}]{SLM98}
\bibinfo{author}{\bibfnamefont{V.}~\bibnamefont{{\v S}pi{\v c}ka}},
  \bibinfo{author}{\bibfnamefont{K.}~\bibnamefont{Morawetz}}, \bibnamefont{and}
  \bibinfo{author}{\bibfnamefont{P.}~\bibnamefont{Lipavsk{\'y}}},
  \bibinfo{journal}{Phys. Rev. E} \textbf{\bibinfo{volume}{64}},
  \bibinfo{pages}{046107} (\bibinfo{year}{2001}).

\bibitem[{\citenamefont{Morawetz
  et~al.}(1999{\natexlab{b}})\citenamefont{Morawetz, {\v S}pi{\v c}ka,
  Lipavsk{\'y}, Kortemeyer, Kuhrts, and Nebauer}}]{MSLKKN99}
\bibinfo{author}{\bibfnamefont{K.}~\bibnamefont{Morawetz}},
  \bibinfo{author}{\bibfnamefont{V.}~\bibnamefont{{\v S}pi{\v c}ka}},
  \bibinfo{author}{\bibfnamefont{P.}~\bibnamefont{Lipavsk{\'y}}},
  \bibinfo{author}{\bibfnamefont{G.}~\bibnamefont{Kortemeyer}},
  \bibinfo{author}{\bibfnamefont{C.}~\bibnamefont{Kuhrts}}, \bibnamefont{and}
  \bibinfo{author}{\bibfnamefont{R.}~\bibnamefont{Nebauer}},
  \bibinfo{journal}{Phys. Rev. Lett.} \textbf{\bibinfo{volume}{82}},
  \bibinfo{pages}{3767} (\bibinfo{year}{1999}{\natexlab{b}}).

\bibitem[{\citenamefont{Morawetz}(2000)}]{MT00}
\bibinfo{author}{\bibfnamefont{K.}~\bibnamefont{Morawetz}},
  \bibinfo{journal}{Phys. Rev. C} \textbf{\bibinfo{volume}{62}},
  \bibinfo{pages}{044606} (\bibinfo{year}{2000}).

\bibitem[{\citenamefont{Morawetz
  et~al.}(2001{\natexlab{b}})\citenamefont{Morawetz, Lipavsk\'y, Normand,
  Cussol, Colin, and Tamain}}]{MLNCCT01}
\bibinfo{author}{\bibfnamefont{K.}~\bibnamefont{Morawetz}},
  \bibinfo{author}{\bibfnamefont{P.}~\bibnamefont{Lipavsk\'y}},
  \bibinfo{author}{\bibfnamefont{J.}~\bibnamefont{Normand}},
  \bibinfo{author}{\bibfnamefont{D.}~\bibnamefont{Cussol}},
  \bibinfo{author}{\bibfnamefont{J.}~\bibnamefont{Colin}}, \bibnamefont{and}
  \bibinfo{author}{\bibfnamefont{B.}~\bibnamefont{Tamain}},
  \bibinfo{journal}{Phys. Rev. C} \textbf{\bibinfo{volume}{63}},
  \bibinfo{pages}{034619} (\bibinfo{year}{2001}{\natexlab{b}}).

\bibitem[{\citenamefont{B{\"a}rwinkel}(1984)}]{Ba84}
\bibinfo{author}{\bibfnamefont{K.}~\bibnamefont{B{\"a}rwinkel}}, in
  \emph{\bibinfo{booktitle}{Proceedings of the 14th International Symposium on
  Rarified Gas Dynamics}} (\bibinfo{publisher}{University of Tokyo Press},
  \bibinfo{address}{Tokyo}, \bibinfo{year}{1984}).

\bibitem[{\citenamefont{Snider}(1991)}]{S91}
\bibinfo{author}{\bibfnamefont{R.~F.} \bibnamefont{Snider}},
  \bibinfo{journal}{J. Stat. Phys.} \textbf{\bibinfo{volume}{63}},
  \bibinfo{pages}{707} (\bibinfo{year}{1991}).

\bibitem[{\citenamefont{March and Santamaria}(1991)}]{MS91}
\bibinfo{author}{\bibfnamefont{N.~H.} \bibnamefont{March}} \bibnamefont{and}
  \bibinfo{author}{\bibfnamefont{R.}~\bibnamefont{Santamaria}},
  \bibinfo{journal}{International Journal of Quantum Chemistry}
  \textbf{\bibinfo{volume}{39}}, \bibinfo{pages}{585} (\bibinfo{year}{1991}).

\bibitem[{\citenamefont{March}(1997)}]{Mar97}
\bibinfo{author}{\bibfnamefont{N.~H.} \bibnamefont{March}},
  \bibinfo{journal}{Phys. Rev. A} \textbf{\bibinfo{volume}{56}},
  \bibinfo{pages}{1025} (\bibinfo{year}{1997}).

\bibitem[{\citenamefont{Morawetz}(2017{\natexlab{a}})}]{M17}
\bibinfo{author}{\bibfnamefont{K.}~\bibnamefont{Morawetz}},
  \bibinfo{journal}{Phys. Rev. E} \textbf{\bibinfo{volume}{96}},
  \bibinfo{pages}{032106} (\bibinfo{year}{2017}{\natexlab{a}}).

\bibitem[{\citenamefont{Morawetz}(2017{\natexlab{b}})}]{M17b}
\bibinfo{author}{\bibfnamefont{K.}~\bibnamefont{Morawetz}},
  \emph{\bibinfo{title}{Interacting systems far from equilibrium - quantum
  kinetic theory}} (\bibinfo{publisher}{Oxford University Press},
  \bibinfo{address}{Oxford}, \bibinfo{year}{2017}{\natexlab{b}}).

\bibitem[{\citenamefont{Chin et~al.}(2010)\citenamefont{Chin, Grimm, Julienne,
  and Tiesinga}}]{Chin10}
\bibinfo{author}{\bibfnamefont{C.}~\bibnamefont{Chin}},
  \bibinfo{author}{\bibfnamefont{R.}~\bibnamefont{Grimm}},
  \bibinfo{author}{\bibfnamefont{P.}~\bibnamefont{Julienne}}, \bibnamefont{and}
  \bibinfo{author}{\bibfnamefont{E.}~\bibnamefont{Tiesinga}},
  \bibinfo{journal}{Rev. Mod. Phys.} \textbf{\bibinfo{volume}{82}},
  \bibinfo{pages}{1225} (\bibinfo{year}{2010}).

\bibitem[{\citenamefont{Morawetz and M{\"a}nnel}(2010)}]{M09}
\bibinfo{author}{\bibfnamefont{K.}~\bibnamefont{Morawetz}} \bibnamefont{and}
  \bibinfo{author}{\bibfnamefont{M.}~\bibnamefont{M{\"a}nnel}},
  \bibinfo{journal}{Phys. Lett. A} \textbf{\bibinfo{volume}{374}},
  \bibinfo{pages}{644} (\bibinfo{year}{2010}).

\bibitem[{\citenamefont{Beth and Uhlenbeck}(1937)}]{BU37}
\bibinfo{author}{\bibfnamefont{G.~E.} \bibnamefont{Beth}} \bibnamefont{and}
  \bibinfo{author}{\bibfnamefont{E.}~\bibnamefont{Uhlenbeck}},
  \bibinfo{journal}{Physica} \textbf{\bibinfo{volume}{4}}, \bibinfo{pages}{915}
  (\bibinfo{year}{1937}).

\end{thebibliography}

\end{document}